\documentclass[aps,prd,onecolumn,amssymb]{revtex4}
\usepackage{graphicx,bm}
\usepackage{amsmath}
\usepackage{amsfonts}
\begin{document}

\newcommand{\be}{\begin{equation}}
\newcommand{\ee}{\end{equation}}
\newcommand{\bea}{\begin{eqnarray}}
\newcommand{\eea}{\end{eqnarray}}
\newcommand{\nn}{\nonumber \\}
\newcommand{\e}{\mathrm{e}}

\title{Phantom Cosmology without Big Rip Singularity}
\author{Artyom~V.~Astashenok$^{1}$,
Shin'ichi~Nojiri$^{2,3}$
Sergei~D.~Odintsov$^{2,4,5}$,
Artyom~V.~Yurov$^{1}$}
\affiliation{$^1$Baltic Federal University of I. Kant, Department of
Theoretical Physics, 236041, 14, Nevsky st., Kaliningrad, Russia \\
$^2$ Department of Physics, Nagoya University, Nagoya
464-8602, Japan \\
$^3$ Kobayashi-Maskawa Institute for the Origin of Particles and
the Universe, Nagoya University, Nagoya 464-8602, Japan \\
$^4$Instituci\`{o} Catalana de Recerca i Estudis Avan\c{c}ats (ICREA)
and Institut de Ciencies de l'Espai (IEEC-CSIC),
Campus UAB, Facultat de Ciencies, Torre C5-Par-2a pl, E-08193 Bellaterra
(Barcelona), Spain \\
$^5$Tomsk State Pedagogical University, Tomsk, Russia
}

\begin{abstract}
We construct phantom energy models with the equation of state parameter $w$
which is less than $-1$, $w<-1$, but finite-time future singularity does not occur.
Such models can be divided into two classes: (i) energy density increases with time
(``phantom energy''  without ``Big Rip'' singularity) and (ii) energy density
tends to constant value with time (``cosmological constant'' with asymptotically
de Sitter evolution). The disintegration of bound structure is confirmed in
Little Rip cosmology. Surprisingly, we find that such disintegration (on
example of Sun-Earth system) may occur even in asymptotically de Sitter phantom
universe consistent with observational data.
We also demonstrate that non-singular phantom models admit wormhole solutions
as well as possibility of Big Trip via wormholes.
\end{abstract}

\maketitle

\noindent {\bf Introduction.}
The discovery of accelerated expansion of the universe \cite{Riess,Perlmutter} 
led to a number of new ideas/solutions in cosmology. Recent
observations of supernovae are consistent with the universe made up 71.3\% of
dark energy and only 27.4\% of a combination of dark matter and baryonic matters
\cite{Kowalski}. Dark energy proposed to explain the cosmic acceleration 
should
  have the strong negative pressure (acting repulsively) in order to
explain the observed accelerating expansion of the universe (for recent
reviews, see \cite{Dark-1,Dark-2,Dark-3,Dark-4,Dark-5,Dark-6}).
The equation of state parameter $w_\mathrm{D}$ for dark energy is negative:
\be
w_\mathrm{D}=p_\mathrm{D}/\rho_\mathrm{D}<0\, ,
\ee
where $\rho_\mathrm{D}$ is the dark energy density and $p_\mathrm{D}$ is
the pressure. We omit subscript $\mathrm{D}$ further for simplicity.

According to the latest cosmological data available, the uncertainties are
still too large to discriminate among the three cases $w < -1$, $w = -1$, and
$w >-1$: $w=-1.04^{+0.09}_{-0.10}$ \cite{PDP,Amman}. If $w<-1$, the
violation of all four energy conditions  occurs. The corresponding phantom
field, which is unstable as quantum field theory \cite{Carrol} but could be
stable in classical cosmology, may be naturally described by a scalar field
with the negative kinetic term. Such Lagrangians  appear in some models of
supergravity \cite{Nilles}, in the gravity theories with higher derivatives
\cite{Pollock} and in string field theory \cite{Aref`eva}.

The additional interest to the models with the phantom fields is caused by
their prediction of a so-called Big Rip singularity
\cite{Starobinsky,Caldwell,Frampton,Nesseris,Diaz,Nojiri}.
Theoretically, the scale factor of the universe becomes infinite at a finite
time in the future which was dubbed  Big Rip singularity. There were proposed
several scenarios to cure the Big Rip singularity:
(i) To consider phantom acceleration as transient phenomenon.
This is possible for a number of scalar potentials.
(ii) To account for quantum effects which may delay/stop the singularity
occurrence \cite{Nojiri-2}.
(iii) To modify the gravitation itself in such a way that it appears to be
observationally-friendly from one side but it cures singularity (for review,
see \cite{review}).
(iv) To couple dark energy with dark matter in the special way \cite{DEM} or
to use special (artificial) form for dark energy equation of state \cite{bamba}.
Note that for quintessence dark energy, other (milder) finite-time singularities
may occur. The corresponding classification of such quintessence-related
finite-time singularities is given in Ref.~\cite{Nojiri-3}. For instance, type
II (sudden) singularity \cite{barrow} or type III singularity \cite{Nojiri-3}
occurs with finite scale factor but infinite energy and/or pressure. Such
quintessence-related finite-time singularities occur for instance, for the
models \cite{Bouhmadi,Khalatnikov,Kamenshchik,Bilic,Bento}
and were also called the ``big freeze'' singularity \cite{Bouhmadi-2,Yurov}.

The closer examination shows that the condition $w<-1$ is not sufficient for a
singularity occurrence. First of all, a transient phantom cosmology is quite
possible. Moreover, one can construct such models that $w$ asymptotically
tends to $-1$ and the energy density increases with time or remains constant but
there is no finite-time future singularity
\cite{Yurov-4,Nojiri-3,barrow,Stefanic,Sahni:2002dx,Frampton-2}.
Of course, most evident case is when Hubble rate
tends to  constant  (cosmological constant or asymptotically de Sitter space),
which may also correspond to the pseudo-rip \cite{Frampton-3}.
Very interesting situation is related with Little Rip cosmology
\cite{Frampton-2} where Hubble rate tends to infinity in the infinite future
(for further investigation, see \cite{Frampton-3,LR,Frampton-4}).
The key point is that if $w$ approaches $-1$ sufficiently rapidly,
then it is possible to have a model in which the time required for singularity
is infinite, i.e., the singularity effectively does not occur. Nevertheless, it
may be shown that even in this case the disintegration of bound structures
takes place in the way similar to Big Rip.

The aim of this article is to develop the method of constructing the
phantom models without finite-time singularity. In Sec. II, a general approach to
this problem is developed. The examples of singular dark energy models are
given there. Sec. III is devoted to the construction of the scalar Little Rip dark energy
models. In Sec. IV, the transient phantom era which ends up at asymptotically
de Sitter universe is investigated.
The corresponding non-singular scalar phantom models are constructed.
It is demonstrated that such models are compatible with latest data from
Supernova Cosmology Project. We show that the dissolution of bound 
structures is
possible in such asymptotically de Sitter universe for special choice of theory
parameters. This gives the observationally-consistent pseudo-rip 
cosmology scenario.
The influence of possible interaction between phantom energy and dark matter
on non-singular cosmological evolution is investigated in Sec. V. In Sec. VI,
the possibilities of wormhole solutions and so-called ``Big Trip'' wormhole
scenario in constructed cosmological models are considered. Some summary and
outlook are given in the Discussion section.

\

\noindent
{\bf Scalar dark energy models with future singularity.}
We start from the FRW equation and the conservation law for spatially flat universe
\be
\label{Fried1}
\left(\frac{\dot a}{a}\right)^2 =  \frac{\rho}{3}\, , \quad
\dot{\rho} =  -3\left(\frac{\dot a}{a}\right)(\rho + p)\, .
\ee
where $\rho$ and $p$ are the total energy density and pressure,
$a$ is the scale factor, $\dot{}=d/dt$,
and we use the natural system units in which $8\pi G=c=1$.

We will examine the future evolution of our universe from the point at which
the pressure and the density are dominated by the dark energy.
For the pressure of dark energy, one can choose the general expression
\be
\label{EoS}
p=-\rho-f(\rho)\, ,
\ee
where $f(\rho)$ is a function of the energy density. The case $f(\rho)>0$ corresponds
to $w<-1$. From (\ref{Fried1}), one can obtain the following
relation between the time coordinate $t$ and $f(\rho)$:
\be
\label{trho}
t = \frac{2}{\sqrt{3}}\int^{x}_{x_{0}} \frac{d x}{f(x)}\, , \quad
x\equiv\sqrt{\rho}\, .
\ee
If the integral (\ref{trho}) converges at $\rho\rightarrow\infty$, we have a
singularity: energy density becomes infinite at a finite future $t=t_\mathrm{f}$ in the
future. The expression for scale factor
\be
\label{arho}
a = a_{0}\exp\left(\frac{2}{3}\int^{x}_{x_{0}} \frac{x d x}{f(x)}\right)\, ,
\ee
indicates that there are two possibilities:

\noindent
(i) scale factor diverges at a finite time (``Big Rip'').

\noindent
(ii) scale factor reaches finite value and singularity
($\rho\rightarrow\infty$) occurs. It
is  type III singularity in the notations of Ref.~\cite{Nojiri-3}.

The simple choice of $f$ corresponding to first scenario is
\be \label{primer}
f(x)=\beta x^{\alpha},\quad 1<\alpha\leq 2\, ,
\ee
where $\beta$ is a positive constant. The scale factor can be written as
\begin{equation}
a=\left\{\begin{array}{ll} A\exp\{(B-Ct)^{-\gamma}\}\, , \quad \alpha\neq 2\, ,
\quad \gamma=\frac{2-\alpha}{\alpha-1} \\
(D-Et)^{-\delta}\, , \quad \alpha=2\, , \quad \delta=2/3\beta\, , \end{array}\right.
\label{sol-sol-sol-sol}
\end{equation}
where $A$, $B$, $C$, $D$, and $E$ are positive constants. The case $\alpha=2$
corresponds to a simplest model of phantom energy with parameter
$w=-1-\beta=\mbox{const}$.

If $\alpha>2$, one has the second possibility: the energy density grows so
rapidly with time that scale factor does not reach the infinite value.

Equivalent description in terms of scalar theory can be derived using the
equations:
\be
\rho=-\dot{\phi}^{2}/2+V(\phi)\, , \quad p=-\dot{\phi}^{2}/2-V(\phi)
\ee
where $\phi$ is a scalar field with potential $V(\phi)$. For the scalar field and
the potential, one can derive following expressions:
\bea
\label{phix}
&& \phi(x)=\phi_{0}\pm\frac{2}{\sqrt{3}}\int_{x_{0}}^{x}\frac{dx}{\sqrt{f(x)}}\, ,\\
\label{Vx}
&& V(x)=x^{2}+f(x)/2\, .
\eea
Combining Eqs.~(\ref{phix}) and (\ref{Vx}), we find the potential as a function of
scalar field. For simplicity we choose the sign ``$+$'' in Eq.~(\ref{phix}), hereafter.
The choice (\ref{primer}) yields for potential (if $\alpha\neq 2$)
\be \label{Vphi}
V(\phi)=F^{\frac{2}{1-\alpha/2}}(\phi-\phi_{0})^\frac{2}{1-\alpha/2}
+ \frac{\beta}{2}F^{\frac{\alpha}{1-\alpha/2}}
(\phi-\phi_{0})^{\frac{\alpha}{1-\alpha/2}}\, ,
\quad F=\sqrt{3\beta}(1-\alpha/2)/2\, .
\ee
For $\alpha=2$, as expected, we have an exponential potential:
\be \label{Vphi2}
V(\phi)=(1+\beta/2)x_{0}^{2}\exp\{\sqrt{3\beta}\left(\phi-\phi_{0}\right)\}\, .
\ee
The key difference between (i) and (ii) for this model is that the potential of
scalar field has a pole in a case of (ii)-singularity.
Note that another type of singularity occurs if $f(x)\rightarrow\infty$ at
$x=x_\mathrm{f}<\infty$, i.e., the pressure of dark energy becomes infinite at finite
energy density. The second derivative of scale factor diverges.

It is interesting however to investigate the phantom energy models without
finite-time future singularities with the help of similar technique.

\

\noindent
{\bf Scalar Little Rip cosmology.}
Let us now consider the models which provide an evolution for the universe
intermediate between de Sitter evolution and
phantom era with the Big Rip. These models were described in detail in
\cite{Frampton-2,Frampton-3,Frampton-4}. The energy density
grows with time but
not rapidly enough for the occurrence of the Big Rip singularity.
According to terminology in
\cite{Frampton-2} we have a so-called ``Little Rip'': eventually a dissolution
of bound structures at some point in the future occurs.

The method of constructing such models is very simple (compare with
\cite{Frampton-4}). In principle one can choose arbitrary monotonic function
$g(x)$ well defined in domain $x>x_{0}$ and satisfying condition
$g(x)\rightarrow\infty$ at $t\rightarrow\infty$. Then one can assume that
\be
f(x)=\frac{1}{g'(x)}\, .
\ee
 From Eqs.~(\ref{trho}) and (\ref{arho}) we have
\bea
&& t=\frac{2}{\sqrt{3}}(g(x)-g(x_{0}))\, ,\nn
&& a=a_{0}\exp\left\{\frac{2}{3}((x g(x)-x_{0}
g(x_{0}))-\frac{2}{3}\int_{x_{0}}^{x}g(x)dx \right\} \, .
\eea
For the scalar field $\phi$ and potential $V(\phi)$,
the following expressions can be written
\be
\phi(x)=\phi_{0}+\frac{2}{\sqrt{3}}\int_{x_{0}}^{x}\sqrt{g'(x)}dx\, ,\quad
V(x)=x^{2}+\frac{1}{2g'(x)}\, .
\ee
The exponential growth of density with time ($g(x)=\ln(x)/\beta$, $\beta>0$)
corresponds to the scalar potential
\be
V(\phi)=\frac{9\beta^{2}}{256}\phi^{4}+\frac{3\beta^{2}}{8}\phi^{2}\, ,\quad
\phi=\phi_{0}\exp\left(\frac{3^{1/2}\beta}{2}t\right)\, ,
\ee
(for simplicity, we put $\phi_{0}=4\sqrt{x_{0}/3\beta}$).
The scale factor $a$ grows with time in accordance with the double exponential law
\be
a=a_{0}\exp\left\{\frac{2x_{0}}{3\beta}\left(\exp\left(
\frac{3^{1/2}\beta t}{2}\right)-1\right)\right\}\, .
\ee
As shown in \cite{Frampton-2}, this model can be matched with the latest data
from Supernova Cosmological Project. The best-fit value for $\beta$ is
$3.46\times10^{-3}$ Gyr$^{-1}$.
Thus, we presented the example of the scalar Little Rip cosmology where the future
singularity does not effectively occurs.

However, it should be noted that the Little Rip produces the disintegration of
bound structures just as in the case of Big Rip. The condition of
disintegration can be derived in the following. The acceleration of the universe
leads to an inertial force on a mass $m$ as seen by a gravitational source
separated by a comoving distance $l$
\be
\label{Fin}
F_\mathrm{in}=ml\frac{\ddot{a}}{a}=ml\frac{4\pi G}{3}(2\rho(a)+\rho'(a)a)\, .
\ee
The structure disintegrates when the inertial force (\ref{Fin}), dominated by
dark energy, becomes equal to the force bounding the
structure. It is convenient to define dimensionless parameter \cite{Frampton-3}
\be
\label{barF}
\bar{F}_\mathrm{in}=\frac{2\rho(a)+\rho'(a)a}{\rho_{0}}\, .
\ee
($\rho_{0}$ is a dark energy density at the present time). The simple
calculations allow to derive the following expression for $\bar{F}_\mathrm{in}$ as
function of time
\be
\bar{F}_\mathrm{in}=2\exp(3^{1/2}\beta t)
+\frac{3\beta}{\rho_{0}^{1/2}}\exp\left(\frac{3^{1/2}\beta t}{2}\right)\, .
\ee
The system of Sun and Earth, for example, disintegrates when $\bar{F}_\mathrm{in}$ reaches
$\sim 10^{23}$.
Therefore, the time required for this event is around $8.5\times 10^3$ Gyr.

\

\noindent
{\bf Phantom models with asymptotically de Sitter evolution.}
Another interesting class of models arises if integral in Eq.~(\ref{trho})
diverges at some finite $x=x_\mathrm{f}<\infty$. The time required for energy density
to reach $\rho=x^{2}_\mathrm{f}$ is infinite, i.e., the expansion of the universe
asymptotically approaches the exponential regime,
which corresponds to the pseudo-rip in \cite{Frampton-3}.
The energy density tends to
the constant value (``cosmological constant'') although the parameter $w$ is
always less than $-1$.

For example, let us assume that
\be \label{AsymDS}
f(x)=A(1-x/x_\mathrm{f})^{\alpha}\, ,
\ee
where $A$ and $\alpha$ are positive constants and we assume $\alpha\geq 1$.
In this case, the integral (\ref{trho}) diverges at $x=x_\mathrm{f}$.
For the case that $\alpha\neq 1, 2$, algebraic calculations allows us
to get the following representation for scale factor:
\be
a(t)=\tilde{a}_{0}\exp(x_\mathrm{f}t/\sqrt{3})\exp(g_{\alpha}(t))\, ,\quad
g_{\alpha}(t)=\frac{2x^2_\mathrm{f}}{3A(2-\alpha)}\left(\frac{A\sqrt{3}(\alpha-1)t}{2x_\mathrm{f}}
+\left(1-\frac{x_{0}}{x_\mathrm{f}}\right)^{1-\alpha}\right)^{1+\frac{1}{1-\alpha}}\, .
\label{as-dS}
\ee
For $1<\alpha<2$ one can easily see that $g(t)\rightarrow 0$ for
$t\rightarrow \infty$.
If $\alpha>2$ $|g(t)|\ll x_\mathrm{f}t/\sqrt{3}$ at $t\rightarrow\infty$.
Therefore the dependence (\ref{AsymDS}) asymptotically tends to de Sitter
solution with vacuum energy density $\rho_{\Lambda}=x^{2}_\mathrm{f}$.

When $t\rightarrow\infty$, the value of scalar field
\be \label{SC1}
\phi=\phi_{0}+\frac{2x_\mathrm{f}}{\sqrt{3A}}\frac{1}{1-\alpha/2}
\left\{\left(1-\frac{x_{0}}{x_\mathrm{f}}\right)^{1-\alpha/2}-\left(\frac{\sqrt{3}
(1-\alpha)At}{x_\mathrm{f}}+\left(1-\frac{x_{0}}{x_\mathrm{f}}\right)^{1-\alpha}
\right)^{\frac{1-\alpha/2}{1-\alpha}}\right\}\, ,
\ee
tends to constant for $1<\alpha<2$ and to
$\phi\rightarrow\pm\infty$ for $\alpha>2$.
As described in the second section, the scalar potential may be found as
\bea
\label{PSC1}
V(\phi) &=& x_\mathrm{f}^{2}\left\{1-\left(\left(1-\frac{x_{0}}{x_\mathrm{f}}
\right)^{1-\alpha/2} - \frac{\sqrt{3A}(1-\alpha/2)(\phi-\phi_{0})}
{2x_\mathrm{f}}\right)^{\frac{1}{1-\alpha/2}}\right\}^{2} \nn
&& +\frac{A}{2}\left\{\left(1-\frac{x_{0}}{x_\mathrm{f}}\right)^{1-\alpha/2}-
\frac{\sqrt{3A}(1-\alpha/2)(\phi-\phi_{0})}{2x_\mathrm{f}}
\right\}^{\frac{\alpha}{1-\alpha/2}}\, .
\eea
The cases $\alpha=1,2$ are more interesting. For $\alpha=1$, the scale factor
behaves as
\be
a(t)=a_{0}\exp(x_\mathrm{f}t/\sqrt{3})\exp(g_{1}(t))\, ,\quad
g_{1}(t)=\frac{2x_\mathrm{f}^{2}}{3A}\left(1-\frac{x_{0}}{x_\mathrm{f}}\right)
(\exp(-\sqrt{3}A t/2 x_\mathrm{f})-1)\, ,
\ee
and the scalar field
\be \label{SC2}
\phi(t)=\phi_{0}+\frac{2 x_\mathrm{f}}{\sqrt{3A}}
\left(1-\frac{x_{0}}{x_\mathrm{f}}\right)^{1/2}(1-\exp(-\sqrt{3}A t/4x_\mathrm{f}))\, ,
\ee
tends asymptotically to maximum (if $x_\mathrm{f}^{2}>A/4$) or minimum (if
$x_\mathrm{f}^{2}<A/4$) of corresponding potential
\bea
\label{PSC2}
V(\phi) &=& x_\mathrm{f}^{2} + \lambda(\phi-\phi^{*})^{4}-\mu^{2}(\phi-\phi^{*})^{2}\, ,\quad
\phi^{*}=\phi_{0} + \frac{2x_\mathrm{f}}{\sqrt{3A}}
\left(1-\frac{x_{0}}{x_\mathrm{f}}\right)^{1/2}\, ,\nn
\lambda &=& \frac{9A^{2}}{16x_\mathrm{f}^{2}},\quad
\mu^{2}=\frac{3A^{2}}{8x_\mathrm{f}^{2}}\left(\frac{4x_\mathrm{f}^{2}}{A}-1\right)\, .
\eea
The choice $\alpha=2$ leads to the exponential potential:
\bea
a(t) &=& a_{0}\exp(x_\mathrm{f}t/\sqrt{3})\exp(g_{2}(t))\, ,\quad
g_{2}(t)=\frac{2x_\mathrm{f}^{2}}{3A}\ln\left(\frac{\sqrt{3}
A(x_\mathrm{f}-x_{0})}{x_\mathrm{f}^{2}}t+1\right)\, , \\
\label{SC3}
\phi &=& \phi_{0}+\frac{2x_\mathrm{f}}{\sqrt{3A}}\ln
\left(1+\frac{\sqrt{3}A}{x_\mathrm{f}^{2}}t\right)\, , \\
\label{PSC3}
V &=& x_\mathrm{f}^{2}-2x_\mathrm{f}\exp(-S)
+\left(1+\frac{A}{2x_\mathrm{f}^{2}}\right)\exp(-2S)\, ,\quad
S=\sqrt{3A}(\phi-\phi_{0})/2x_\mathrm{f}\, .
\eea
The appearance of exponential potentials may indicate to some connection with
string theory. Eqs.~(\ref{SC3}) and (\ref{PSC3}) show that in the infinite
future, the scalar field also goes to infinity and the scalar field climbs up
the potential to constant $V\rightarrow x^{2}_\mathrm{f}$.

The above example is a good theoretical illustration of the new phantom
energy models mimicking vacuum energy. The dark energy with such a behavior can
be realized if the function $f(x)$ is equal to zero at $x=x_\mathrm{f}$ and
the integral (\ref{trho}) diverges at $x=x_\mathrm{f}$.
In the vicinity of this point, the arbitrary
function $f$ satisfying these conditions can be expanded as
\[
f(x)=(x-x_\mathrm{f})^{\alpha}+\mbox{\textit{O}}((x-x_\mathrm{f})^{\alpha})\, ,
\quad \alpha\geq 1\, .
\]
Thus for $t\rightarrow\infty$, one can expect that Eqs.~(\ref{SC1}),
(\ref{SC2}), and (\ref{SC3}) for $\phi(t)$ and (\ref{PSC1}), (\ref{PSC2}),
and (\ref{PSC3}) for $V(\phi)$ will be satisfied.
The case $\alpha=1$ corresponds to the most rapid growth of the energy
density with time in which any singularity does not occur.

Let us consider another model:
\be
f(x)=A\cos^{2}\left(\frac{\pi x}{2x_\mathrm{f}}\right).
\ee
In the vicinity $x=x_\mathrm{f}$, we find
$f(x)\approx\frac{A\pi^2}{4x^{2}_\mathrm{f}}(1-x/x_\mathrm{f})^{2}$
and we can conclude that the potential of scalar field looks like that
in (\ref{PSC3}) at $t\rightarrow\infty$.
Indeed, this is correct. For such a model, one gets
\be
V(\phi)=x_\mathrm{f}^{2}\left\{\frac{2}{\pi}\arctan\exp(S)\right\}
+\frac{2A\exp(S)}{(1+\exp(S))^{2}}\, ,\quad
S=\frac{\sqrt{3A}\pi}{x_\mathrm{f}}(\phi-\phi_{0})\, .
\ee
The interesting question is: could such models in principle describe latest
supernova data from the Supernova Cosmology Project? The analysis shows that it
is possible. Moreover the construction of such models is trivial. For example
let us choose
\be \label{Toy}
f(x)=\beta x^{1/2} (1-(x/x_\mathrm{f})^{3/2}) \, ,
\ee
and assume that dark energy density varies from $0$ to
$\rho_\mathrm{f}=x_\mathrm{f}^{2}$.
Let $t=t_{0}$ and $\rho=\rho_{0}$ at current universe.
Eq.~(\ref{arho}) allows to write the following relation between the dark energy
density $\rho$ and the redshift $z=a_{0}/a-1$:
\be
\label{Toy2}
\rho(z)=\rho_\mathrm{f}\left(1-(1+z)^{\gamma}(1-\Delta)\right)^{4/3}\, ,\quad
\gamma=3\beta \rho_\mathrm{f}^{-3/4}/2\, ,\quad
\Delta=\left(\frac{\rho_{0}}{\rho_\mathrm{f}}\right)^{3/4}\, .
\ee
The equation of state parameter $w_{0}$ is
\be
w_{0}=-1-\frac{2\gamma}{3}\frac{1-\Delta}{\Delta}\, .
\ee
For the scale factor, we have the parametric expression
\bea
a(v) &=& \frac{a_{0}}{(1-v^{3/2})^{2/3\gamma}} \, ,\\
\label{tv}
t(v) &=& t_{0}+\frac{2\sqrt{3}}{\gamma x_\mathrm{f}}
\left(\frac{1}{6}\ln(v^2+v+1)
+\frac{1}{\sqrt{3}}\arctan\frac{2v+1}{\sqrt{3}}-\frac{\pi}{6\sqrt{3}}
  -\frac{1}{3}\ln(1-v)\right)\, .
\eea
The parameter $v$ varies from $0$ (at $t=t_{0}$) to $1$ (at $t\rightarrow\infty$).
The last term in Eq.~(\ref{tv}) dominates at $v\rightarrow 1$. In this case, with
a good accuracy one can write
\be
a=\left(\frac{2}{3}\right)^{2/3\gamma}a_{0}
\exp\left(\frac{x_\mathrm{f}(t-t_{0})}{\sqrt{3}}\right)\, .
\ee
For the dark and baryonic matter densities,
\be
\rho_{m}=\rho_{m0}(1+z)^{3}\, .
\ee
Therefore, the dependence of luminosity distance $D_\mathrm{L}$ from redshift $z$ for
this model is given by
\bea
\label{DL}
D_\mathrm{L} &=& \frac{c}{H_{0}}(1+z)\int_{0}^{z}\left(\Omega_{m}
(1+z)^{3}+\Omega_\mathrm{D}h(z)\right)^{-1/2}d z\, ,\nn
h(z) &=& \Delta^{-4/3}(1-(1+z)^{\gamma}(1-\Delta))^{4/3}\, .
\eea
In Eq.~(\ref{DL}), $\Omega_{m}$ and $\Omega_\mathrm{D}$ express the fractions
of matters and dark energy in the total energy budget correspondingly.
For the standard $\Lambda$CDM cosmology, as well-known, we find
\be \label{DLSC}
D^{SC}_\mathrm{L}=\frac{c}{H_{0}}(1+z)\int_{0}^{z}
\left(\Omega_{m}(1+z)^{3}+\Omega_{\Lambda}\right)^{-1/2}d z\, .
\ee
If $\gamma$ and $\Delta$ are very close to zero and $1$ correspondingly
($\beta$ is small and $\rho\rightarrow\rho_\mathrm{f}$), then $h(z)$ is close to 1.
Therefore the functions (\ref{DL}) and (\ref{DLSC}) are essentially
indistinguishable (especially in the observable range $0<z<1.5$).
The equation of state parameter is $w\approx -1$.
\begin{figure}
\includegraphics[scale=1]{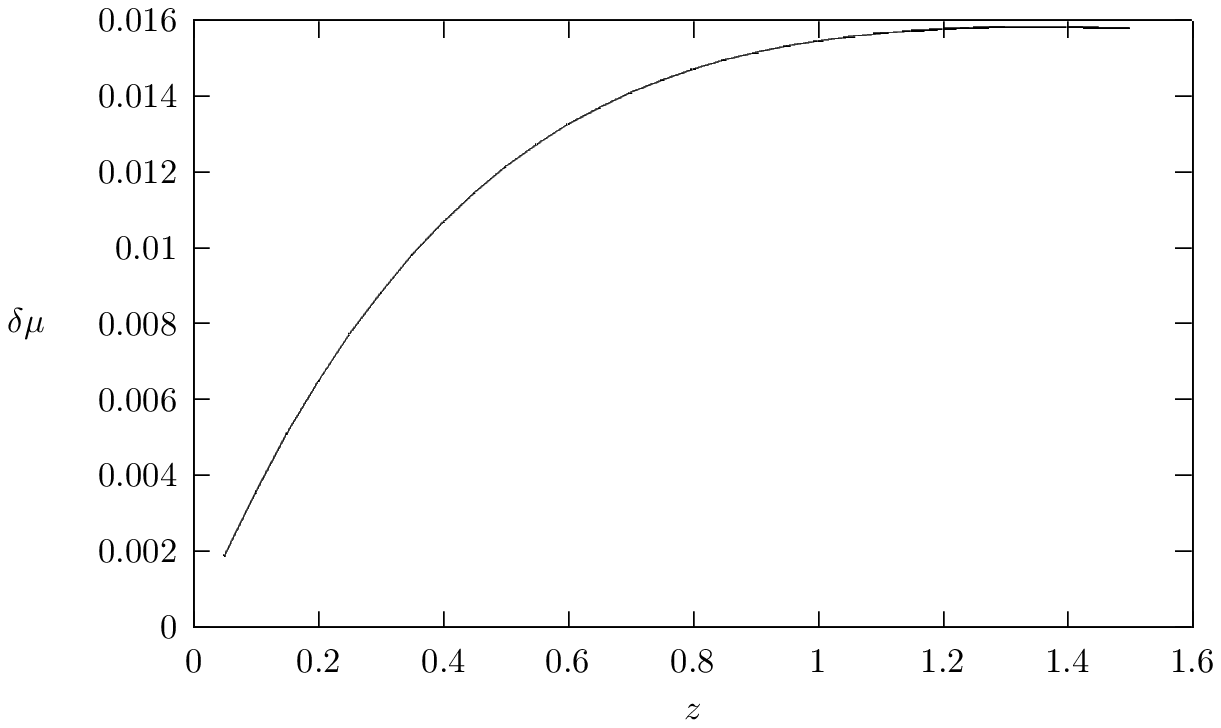}\\
\caption{The difference between modulus extracted from the toy dark energy model
(\ref{Toy}) and from the standard $\Lambda$CDM-model.}\label{1}
\end{figure}
Another choice of parameters
\[
\Delta=0.5,\quad \gamma=0.075
\]
corresponds to current equation of state parameter $w_{0}=-1.05$. The
difference $\delta \mu=5\lg (D/D^{SC})$ ($\mu$ is a distance modulus) for
$0<z<1.5$ is depicted on Fig.~\ref{1} and does not exceed $0.016$ (for
$\Omega_{m}=0.28$, $\Omega_{\Lambda}=\Omega_\mathrm{D}=0.72$).
Taking into account that the errors in definition of SNe modulus are
$\sim 0.075\div 0.5$, we conclude that our model fits these data with
excellent precision.

\begin{figure}
\includegraphics[scale=1]{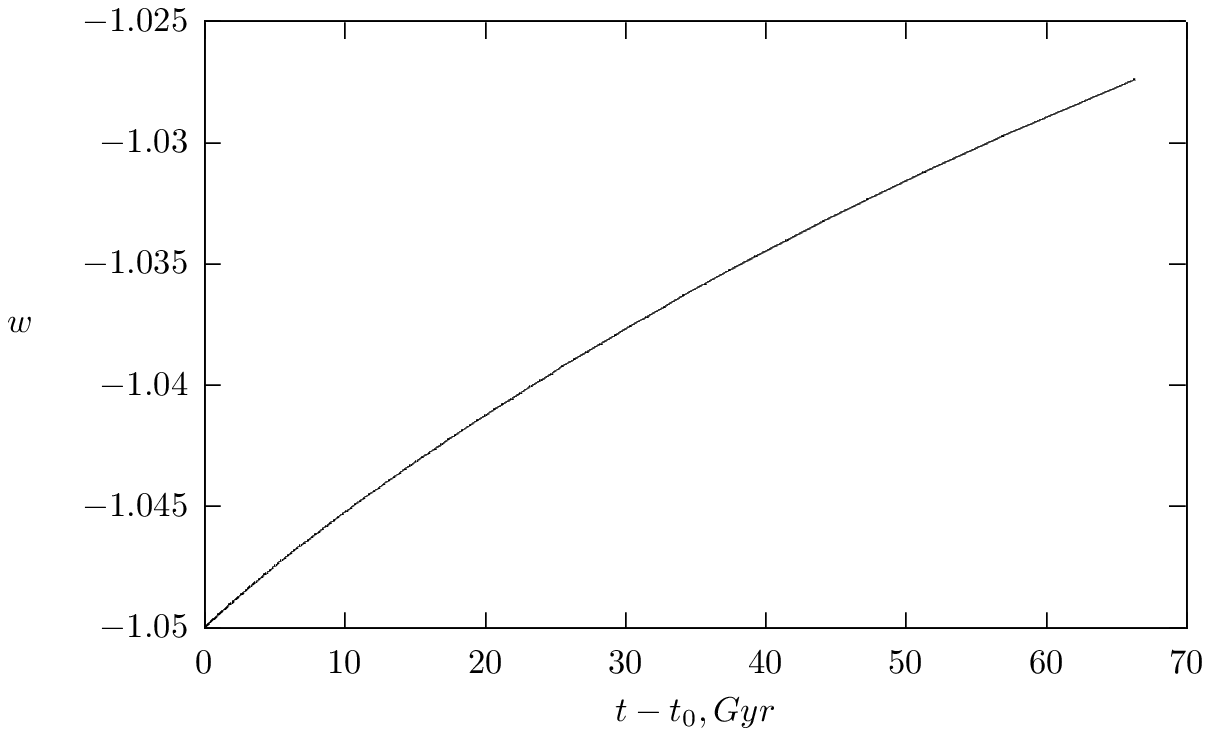}\\
\caption{The time dependence of $w$ for the model (\ref{Toy}).}\label{2}
\end{figure}

\begin{figure}
\includegraphics[scale=1]{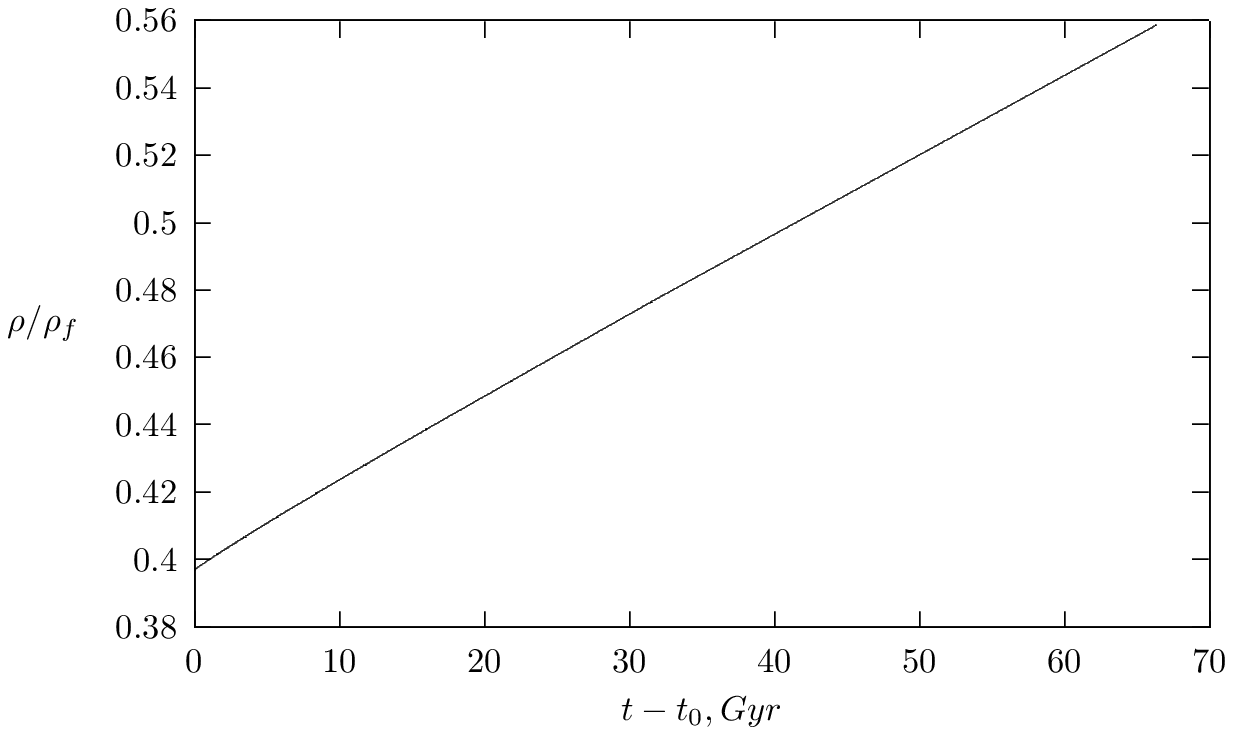}\\
\caption{The time dependence of the dark energy density for the model
(\ref{Toy}).}\label{3}
\end{figure}

The equation of state parameter $w$ and phantom energy density very slowly
increase with time (Fig.~\ref{2}, Fig.~\ref{3}).


Thus, we presented non-singular phantom dark energy which evolves to
asymptotically de Sitter space and satisfies the observational bounds.

Let us  compare the model (\ref{Toy}) with observational data in more
detail.
One defines deceleration parameter $q_0$, jerk parameter $j_0$ as follows
\bea
\label{DDD1}
&& q_0 = - \left. \frac{1}{a H^2} \frac{d^2 a}{dt^2}\right|_{t=t_0}
= - \left. \frac{1}{H^2} \left\{ \frac{1}{2} \frac{d \left( H^2 \right)}{dN}
+ H^2 \right\} \right|_{N=0} \, ,\nn
&& j_0 = \left. \left\{ \frac{1}{a H^3} \frac{d^3 a}{dt^3} \right|_{t=0}
= \left. \frac{1}{2H^2} \frac{d^2 \left( H^2 \right)}{dN^2}
+ \frac{3}{2H^2} \frac{d \left( H^2 \right)}{dN} + 1 \right\} \right|_{N=0} \, .
\eea
Here $N$ is the e-foldings defined by
\be
\label{e-foldings}
N \equiv - \ln \left(1+z\right)\, .
\ee
For the current universe $t=t_0$, we have $N=0$.
Since
\be
\label{Toy3}
H^2 = \frac{1}{3} \left(\rho + \rho_m\right)
= \frac{1}{3} \left( \rho_f \left( 1 - \e^{-\gamma N} \left(1 - \Delta \right) \right)^{4/3}
+ \rho_{m0} \e^{-3N}\right) \, ,
\ee
one gets
\bea
\label{DDD1B}
q_0 &=& \frac{2}{3}\gamma \left(\Delta^{-1} - 1 \right) \Omega_\Lambda
+ \frac{3}{2} \Omega_m - 1\, ,\nn
j_0 &=& \left( \frac{2}{9}\gamma^2 \left(\Delta^{-1} - 1 \right)^2
+ \left( - \frac{2}{3}\gamma^2 + 2 \gamma \right) \left(\Delta^{-1} - 1 \right)
\right) \Omega_\Lambda + 1 \, .
\eea
Since $\Delta=0.5$, $\gamma=0.075$, $\Omega_{m}=0.28$,
and $\Omega_{\Lambda}=0.72$, we find
\be
\label{Toy4}
q_0 = - 0.54\, ,\quad j_0 = 1.11\, .
\ee
In case of $\Lambda$CDM model, which corresponds to $\Delta=0$,
we have $q_0=- 0.58$ and $j_0=1$, which is not so different from
the values  (\ref{Toy4}).

Just as it was done in previous section, we can estimate the possibility of
the disintegration of bound system in our model (\ref{Toy}). It is easy to derive
the following parametric representation for the dimensionless inertial force
$\bar{F}_\mathrm{in}$ in (\ref{barF}) as follows,
\bea
&& \bar{F}_\mathrm{in}(u) = 2\Delta^{-4/3}(1-(1+u)^{\gamma}(1-\Delta))^{1/3}
\left\{1+\left(\frac{2}{3}\gamma-1\right)(1+u)^{\gamma}(1-\Delta)\right\}\, , \\
&& t-t_{0}=\frac{1}{H_{0}}\int_{u}^{0}du(1+u)^{-1}
\left(\Omega_{m}(1+u)^{3}+\Omega_\mathrm{D}h(u)\right)^{-1/2}\, .
\eea
The variable $u=a_{0}/a-1$ varies from $0$ (at present time) to $-1$
(when $t\rightarrow\infty$). The function $h(u)$ coincides with $h(z)$ in
Eq.~(\ref{DL}) by changing $z\rightarrow u$.
The inertial force asymptotically tends to $2\Delta^{-4/3}$.
Therefore, the disintegration of the system of Sun and Earth
can occur if $\Delta\leq \Delta_{min}=10^{-17}$.
The parameter $\gamma$ at
$\Delta=\Delta_{min}$ may vary from $0$ (when $w_{0}=-1$) to $2\times 10^{-18}$
(when $w_{0}=-1.14$ -- the lower bound from observations).
The analysis shows that for any $\Delta<\Delta_{min}$ and $0<\gamma<10^(-1)\Delta$,
our model also describes the SNe data.
Therefore, the current observations in principle do not contradict
with the possible disintegration of bound system in the models with
an asymptotically de Sitter expansion. Note that such models were dubbed 
pseudo-rip in Ref.~\cite{Frampton-3}.

One remark is in order. The asymptotically de Sitter expansion of the universe
can occur not only in the models with the phantom energy but also for
the quintessence dark energy ($-1<w<0$).
For the quintessence, the function $f(x)$ in (\ref{EoS}) is negative and
the energy density decreases with time.
Eqs.~(\ref{trho}), (\ref{arho}), and (\ref{Vx}) do not change, for Eq.~(\ref{phix}),
it is sufficient to replace $f(x)\rightarrow - f(x)$.
One can consider our model (\ref{AsymDS}) and assume
that $\alpha=1$ and $x_{0}>x_\mathrm{f}$.
The energy density asymptotically tends to $\rho\rightarrow x^{2}_\mathrm{f}$ but
the equation of state parameter to $w>-1$.
Therefore one can conclude that the point $x=x_\mathrm{f}$ is attractor
in the case $f(x)=A(1-x/x_\mathrm{f})$.

\

\noindent
{\bf Coupled phantom models.}
Realistic cosmological scenarios should take into account that the dark energy does
not have a single component of universe energy. It is quite appealing to include the
possible interaction between dark energy and dark matter (for recent
discussion, see \cite{Wei,Chimento,Cai}). We shall see that the addition of such a
coupling may lead to interesting effects in the non-singular phantom cosmology.
The cosmological solutions with interacting phantom energy exhibit much richer
behavior than those considered above.

It is customary to assume that phantom dark energy and dark matter interact
through a coupling term $Q$ as
\bea
\label{Inter1}
&& \dot{\rho}_{m}+3H\rho_{m}=-Q\, , \\
\label{Inter2}
&& \dot{\rho}_\mathrm{D}+3H(\rho_D+p_D)=Q\, .
\eea
Let us introduce variables $x$ and $y$
\[
\rho_\mathrm{D}=x^{2},\quad \rho_{m}=y^{2}\, .
\]
For convenience, we assume that
\be
Q=3Hq(x,y)\, ,
\ee
where $q(x,y)$ is an arbitrary function of variables $x$ and $y$.
Then the Hubble rate $H$ is given by
\be
H=\left(\frac{x^2+y^{2}}{3}\right)^{1/2}\, ,
\ee
and Eqs.~(\ref{Inter1}) and (\ref{Inter2}) can be rewritten as
\bea
\label{Sys1}
\dot{y} &=& -\frac{3^{1/2}(x^2+y^2)^{1/2}}{2y}\left(y^2+q(x,y)\right) \, ,\\
\label{Sys2}
\dot{x} &=& \frac{3^{1/2}(x^2+y^2)^{1/2}}{2x}\left(f(x)+q(x,y)\right) \, .
\eea
We can obtain critical points of the system (\ref{Sys1}), (\ref{Sys2})
satisfying following conditions, $\dot{x}_\mathrm{f}=0$ and $\dot{y}_\mathrm{f}=0$.
Note that these critical points  must satisfy the requirement $x_\mathrm{f}\geq0$,
$y_\mathrm{f}\geq0$.
Unfortunately, there are no well-motivated direct observational or theoretical
bounds on the type of interaction.
Usually, simple interactions $q\sim \rho_{m}$, $q\sim \rho_\mathrm{D}$, and
$q\sim (\rho_\mathrm{D}+\rho_{m})$ are extensively investigated.

For simplicity, we restrict ourselves to the cases $q(x,y)\equiv g(x)$, i.e.,
the intensity of interaction depends only from the phantom energy density and
the rate of the Hubble rate $H$. Let us consider two examples,

\noindent
1. The simplest phantom energy model with $f(x)=\beta x^{2}$ and $w=-1-\beta$.
Without interaction, one gets the Big Rip singularity. One can assume that
\[
q(x,y)=-\alpha x^4,\quad \alpha>0\, ,
\]
i.e., interaction leads to transformation of phantom energy into dark matter.
On the other hand, the density of dark energy grows with time. Eventually the
dynamical equilibrium between these processes is established. The equilibrium
phantom energy density is
\[
\rho^{eq}_\mathrm{D}=\frac{\beta}{\alpha}\, .
\]
The density of dark matter otherwise decreases as universe expands but the dark
matter born out from the phantom energy. The density of the dark matter
asymptotically tends to
\[
\rho^{eq}_{m}=\frac{\beta^{2}}{\alpha}\, .
\]
Then the Hubble rate tends to
\[
H\rightarrow \left(\frac{\beta+\beta^{2}}{3\alpha}\right)^{1/2}\, ,
\quad t\rightarrow\infty\, ,
\]
i.e., we have expansion according to the de Sitter law with ``effective''
cosmological constant
\[
\Lambda^\mathrm{eff}=\frac{\beta}{\alpha}(\beta+1)\, .
\]
In other words, coupling with dark matter changes the qualitative behavior of
phantom dark energy making it to be non-singular, in the way similar to the one
described in Ref.~\cite{DEM}.

\noindent
2. The ``switch'' from the Little Rip to the asymptotically de Sitter expansion.
As an example, let us choose
\[
f(x)=\beta x,\quad q(x)=-\alpha x^{2}, \alpha>0\, .
\]
The interpretation of interaction is the same as in the previous model. Instead of
the Little Rip expansion, the quasi-de Sitter expansion occurs with the effective
cosmological constant:
\[
\Lambda^\mathrm{eff}=\frac{\beta}{\alpha^{2}}(\alpha+1)\, .
\]
Thus, we demonstrated that coupling of phantom dark energy with dark matter may
help in the transition from singular accelerating expansion
to the non-singular one. Note that stability of such cosmologies may be
investigated in the same way as in Ref.~\cite{ito} where it is shown that the Little
Rip often may be stable if we compare it with de Sitter universe.

\

\noindent
{\bf ``Big Trip'' in phantom cosmology without singularities.}
One of the very interesting points in phantom cosmology is connected with
wormholes.  First of all, the existence of the static, spherically symmetric
wormhole solutions of the gravitational field equations in the absence of ghost (or
phantom) degrees of freedom is impossible, as it was shown in
\cite{Kiril-Starobinsky}. We discuss here wormholes application in phantom
cosmology under discussion.

It was shown in \cite{BigTrip-04,BigTrip-06} that as one goes towards
the Big Rip, there would occur the process of the fast wormhole swelling taking
the size of the wormhole throat to infinity during the finite time.
The reason of such a striking behavior is the phantom
energy accretion onto wormhole. This accretion induces an increase of the
wormhole throat radius so quick that the wormhole would engulf the
entire universe before this reached the Big Rip. Such a result has been dubbed as
``Big Trip'' and was later criticized in Ref.~\cite{BigTrip-Faraoni}.
The rejoinder was contained in \cite{BigTrip-rejoinder-1}.
The issue remains open up to now and it is not proposed  to deal at length
with this discussion.
All we need here is to consider the formal possibility of the Big Trip in phantom
models without Big Rip.  As we shall see, the Big Trip in phantom cosmology is a
common occurrence, that often happens even if we consider models with
asymptotic de Sitter evolution.

The equation which describes evolution of the throat radius of a Morris-Thorne
wormhole $b=b(t)$ due to dark energy accretion has the form
\begin{equation}
{\dot b}=-2\pi^2D b^2(1+w)\rho\, ,
\label{dotb}
\end{equation}
where $D$ is positive dimensionless constant and $\rho$ is the energy density
of the dark energy fluid. This equation is immediate consequence of the result
for the dynamics of the mass of a black hole due to fluid accretion
\cite{Bab-Doc-Ero}. In this section we consider four examples of phantom models
without  finite-time singularity.

\noindent
1. The first and second model was obtained in
\cite{Frampton-2,Frampton-4}. Let put the Hubble rate as
$H=H_0\mathrm{e}^{\lambda t}$ with positive constants $H_0$ and $\lambda$.
As a result we
have the scale factor, the density and the pressure in the form
\begin{equation}
\begin{array}{l}
\displaystyle{
a(t)=a_0\exp\left[ \frac{H_0}{\lambda}\left(\mathrm{e}^{\lambda t}-1\right)
\right]\, ,}\\
\\
\displaystyle{
\rho(t)=H_0^2\mathrm{e}^{2\lambda t},\qquad p(t)=-\frac{H_0}{3}\mathrm{e}^{\lambda
t}\left[2\lambda+3H_0\mathrm{e}^{\lambda t}\right]}\, ,
\end{array}
\label{Sol-1}
\end{equation}
so
\[
\displaystyle{
w(t)=-1-\frac{2\lambda}{3H_0} \mathrm{e}^{-\lambda t}}\, ,
\]
and at $t\to\infty$ one get the asymptotic de Sitter universe.

Substituting (\ref{Sol-1}) into (\ref{dotb}), one obtains
\begin{equation}
\displaystyle{
b(t)=\frac{3}{4\pi^2 DH_0\left(\mathrm{e}^{\lambda t_{_{BT}}}
  - \mathrm{e}^{\lambda  t}\right)}}\, .
\label{b-1}
\end{equation}
Thus we have the Big Trip at
\[
\displaystyle{
t=t_{_{BT}}=\frac{1}{\lambda}\log\left(b_0+\frac{3}{4\pi^2 DH_0}\right)}\, ,
\]
where $b_0$ is the throat radius of the wormhole at $t=0$: $b=b(t)$.

\noindent
2. Let $H=H_0-H_1\mathrm{e}^{-\lambda t}$. This is the most clear example of universe
with de Sitter asymptotic behavior, if $H_0>0$ and $\lambda>0$. After integration of
FRW equations we have
\[
\displaystyle{
w(t)=-1-\frac{2\lambda H_1\mathrm{e}^{-\lambda t}}{2\left(H_0-H_1
\mathrm{e}^{-\lambda t}\right)^2}\, ,}
\]
and $w(t)\to -1$ at $t\to +\infty$. It is interesting to note that this
solution contains so called w-singularity \cite{w-singularity}, to be more
precise, some generalization of w-singularity which was obtained in
\cite{Brane-like-Yu}. In fact, at
\[
t=t_w=-\frac{1}{\lambda}\log\frac{H_0}{H_1}\, ,
\]
$w(t_w)=-\infty$ although $\rho(t_w)=0$ and $p(t_w)=-2\lambda H_0/3\ne \infty$.
Sure, the w-singularity occurs if both $H_0$ and $H_1$ are positive constants.
If $H_1<0$ then $w(t)$ is always finite.

After  calculations we have  the throat radius of the wormhole in the form:
\begin{equation}
\displaystyle{
b(t)=\frac{3b_0}{3-4\pi^2DH_1b_0\left(1-\mathrm{e}^{-\lambda t}\right)}\, .}
\label{b-2}
\end{equation}
For negative values of $H_1$ we have Big Trip for any values of positive
parameters $b_0$ and $D$ but have no w-singularity. If $H_1>0$ then Big Trip
takes place only for wormholes with
\[
b_0>\frac{3}{4\pi^2DH_1}\, ,
\]
at
\[
\displaystyle{
t=t_{_{BT}}=-\frac{1}{\lambda}\log\left(1-\frac{3}{4\pi^2 D H_1b_0}\right)\, .}
\]
For positive $H_1$ we have w-singularity and
\[
b(t_w)=\frac{3b_0}{3+4\pi^2Db_0(H_0-H_1)}\, .
\]
If
\[
H_1>H_0+\frac{3}{4\pi^2b_0}\, ,
\]
then Big Trip takes place after w-singularity, otherwise - before. In any case,
it is clear that w-singularity does not influence the evolution of wormhole at
all, since it is not real physical singularity.

\noindent
3. Now let consider the solution (\ref{sol-sol-sol-sol}) with $\alpha\ne 2$.
Since we are interested in solutions without Big Rip one should take
$\gamma=-g<0$; this  is possible if $\alpha<1$ ($g>1$) and $\alpha>2$
($0<g<1$).  We have phantom model  $w<-1$ for the $\alpha<1$, and $w>-1$ for
the $\alpha>2$. It is interesting to note that if $g>1$ ($\alpha<1$, $w<-1$)
then for $t\to t_s=B/C$ we have $\rho\to 0$. If $1<g<2$ ($\alpha<0$, $w<-1$)
then at $t\to t_s$ $p\to -\infty$, $\rho\to 0$. This is type II singularity
\cite{Nojiri-3}. If $g>2$ ($0<\alpha<1$, $w<-1$) then $\rho(t_s)=p(t_s)=0$,
$w(t_s)=-\infty$. This is exact w-singularity for the phantom case. For $g=2$
($\alpha=0$) we have generalized w-singularity.

The throat radius of the wormhole is
\[
\displaystyle{
b(t)=\frac{3b_0}{3-4\pi^2DgCb_0\left(B^{g-1}-(B-Ct)^{g-1}\right)}\, .}
\]
so the Big Trip takes place if and only if $g>1$, i.e., $w<-1$ which is the case
for the $\alpha<1$, as expected.

\noindent
4. At last we consider the solution (\ref{as-dS}) with $g<0$. Here we have the
generalized $w$-singularity ($\rho(t_w)=0$, $p(t_w)=-A/3$), at
\[
\displaystyle{
t_w=\frac{2\sqrt{3}\left(|g|+1\right)
x_f}{3A}\left[1-\left(1-\frac{x_0}{x_f}\right)^{-1/(|g|+1)}\right]\, ,}
\]
and the Big Trip at
\[
\displaystyle{
t_{_{BT}}=\frac{2x_f(|g|+1)}{A\sqrt{3}}\left(\left(1-\frac{x_0-\delta}{x_f}
\right)^{-1/(|g|+1)}-\left(1-\frac{x_0}{x_f}\right)^{-1/(|g|+1)}\right)\, ,}
\]
with $\delta=3\sqrt{3}/(4\pi Db_0)>0$.

Thus, smooth exit from the phantom inflationary phase can still be tentatively
recovered by considering a Big Trip scenario where the primordial phantom
universe would travel in time towards a future universe (filled with, for
example, usual radiation, see \cite{BigTrip-04}). Such ``exit from inflation''
is possible in  phantom models both with and without the future Big Rip
singularity.

\

\noindent
{\bf Conclusion.}
In summary,
the relatively simple method for constructing phantom energy models without
finite-time future singularity is developed. The dark energy models without future
singularity are attractive from the physical viewpoint because the occurrence of
finite-time singularity  may lead to some inconsistencies. The equivalent
description of the Little Rip cosmology where singularity effectively disappears,
via fluid or scalar-tensor theory is presented.
Phantom models with asymptotically de Sitter evolution are described.
It is demonstrated that asymptotically de Sitter expansion can be realized in
the class of exponential or power-law scalar potentials. Generalization for
phantom models coupled with dark matter is also discussed.
It is interesting to note that disintegration of bound structure (on the
example of the system of Sun and Earth) in some asymptotically de Sitter phantom universe
may occur for observationally acceptable choice of parameters.

We have shown that current data make it essentially
impossible to determine whether or not the universe will end in a future
singularity. The above scalar dark energy models represent natural alternative
for $\Lambda$CDM model, which also leads to non-singular cosmology.
Nevertheless, even for the non-singular asymptotically de Sitter universe, the
possibility of dramatic rip which may lead to the disappearance of bound structures
in the universe remains to be possible.

It is confirmed that phantom models without the Big Rip may lead to wormholes
solutions.
It is demonstrated that the possible Big Trip in the phantom cosmology can
happen even if we consider the models with asymptotic de Sitter evolution.

The presence of numbers of free possible parameters (for instance, the choice of
equation of state (\ref{EoS})) gives enough space for fine-tuning the models
which can be useful for fitting with observational data. Hence, the described
method may be very useful for confronting of theoretical models with coming
observational data.

\

\noindent
{\bf Acknowledgments.}
We are grateful to P.~Frampton and R.~Scherrer for very useful discussions.
S.N. is supported by Global COE Program of Nagoya University (G07)
provided by the Ministry of Education, Culture, Sports, Science \&
Technology and by the JSPS Grant-in-Aid for Scientific Research (S) \# 22224003
and (C) \# 23540296.
The work by SDO has been supported in part by MICINN (Spain) project FIS2010-15640,
by AGAUR 2009SGR-994, by JSPS Visitor Program S11135 (Japan), by LRSS-224.2012.2 
(Russia), and by BFU (Russia).


\begin{thebibliography}{99}

\bibitem{Riess} A.~G.~Riess et al., Astron.\ J. {\bf 116}, 1009 (1998).

\bibitem{Perlmutter} S.~Perlmutter et al., Ap.\ J. {\bf 517}, 565 (1999).

\bibitem{Kowalski} M.~Kowalski, Ap.\ J.\ {\bf 686}, 74 (2008).

\bibitem{Dark-1} E.~Copeland, M.~Sami and S.~Tsujikawa, Int.\ J.\ Mod.\ Phys.
{\bf D 15}, 1753 (2006).

\bibitem{Dark-2} R.~Caldwell and M.~Kamionkowski,
Ann.\ Rev.\ Nucl.\ Part.\ Sci. {\bf 59}, 397 (2009).

\bibitem{Dark-3} R.~Durrer and R.~Maartens, Gen.\ Rel.\ Grav. {\bf 40}, 301 (2008).

\bibitem{Dark-4} J.~Frieman and M.~Turner,
Ann.\ Rev.\ Astron.\ Astrophys. {\bf 46}, 385 (2008).

\bibitem{Dark-5} A.~Silvestri and M.~Trodden,
Rept.\ Prog.\ Phys., {\bf 72}, 096901 (2009).

\bibitem{Dark-6} M.~Li, X.~Li, S.~Wang and Y.~Wang,
Commun.\ Theor.\ Phys. {\bf 56}, 525 (2011).

\bibitem{PDP} Review of Particle Physics, 102 (2010).

\bibitem{Amman} R.~Amanullah et al., Ap.\ J. {\bf 716}, 712 (2010).

\bibitem{Carrol} S.~M.~Carroll, M.~Hofman and M.~Trodden,
Phys.\ Rev. {\bf D68}, 023509 (2003).

\bibitem{Nilles} H.-P.~Nilles, Phys.\ Rep. {\bf 110}, 1 (1984).

\bibitem{Pollock} M.~D.~Pollock, Phys.\ Lett. {\bf B 215}, 635 (1988).

\bibitem{Aref`eva} I.~Ya.~Aref'eva, S.~Yu.~Vernov and A.~S.~Koshelev,
Theor.\ and\ Math.\ Phys. {\bf 148}, 23 (2006).

\bibitem{Starobinsky} A.~A.~Starobinsky, Grav.\ Cosmol. {\bf 6}, 157 (2000).

\bibitem{Caldwell} R.~R.~Caldwell, Phys.\ Lett. {\bf B 545} 23 (2002); \\
R.~R.~Caldwell, M.~Kamionkowski and N.~N.~Weinberg,
Phys.\ Rev.\ Lett. {\bf 91}, 071301 (2003).

\bibitem{Frampton} P.~H.~Frampton and T.~Takahashi,
Phys.\ Lett. B {\bf 557}, 135 (2003).

\bibitem{Nesseris} S.~Nesseris and L.~Perivolaropoulos,
Phys.\ Rev. {\bf D 70}, 123529 (2004); \\
V.~Faraoni, Class.\ Quant.\ Grav. {\bf 22}, 3235 (2005); \\
B.~McInnes, Nucl.\ Phys.\ {\bf B 718}, 55 (2005).

\bibitem{Diaz} P.~F.~Gonz$\acute{a}$lez-Di$\acute{a}$z,
Phys.\ Lett.\ B {\bf 586}, 1 (2004); Phys. Rev. {\bf D 69}, 063522 (2004).

\bibitem{Nojiri} S.~Nojiri and S.~D.~Odintsov,
Phys.\ Rev.\ {\bf D 70}, 103522
(2004); Phys. Lett. {\bf B 562}, 147 (2003).

\bibitem{Nojiri-2} E.~Elizalde, S.~Nojiri and S.~D.~Odintsov,
Phys.\ Rev.\ {\bf D 70}, 043539 (2004); \\
S.~Nojiri and S.~D.~Odintsov, Phys. Lett. {\bf B 595}, 1 (2004).

\bibitem{review} S.~Nojiri and S.~D.~Odintsov,
Phys.\ Rept. {\bf 505}, 59(2011), arXiv:1011.0544.

\bibitem{DEM} S.~Nojiri and S.~D.~Odintsov,
Phys.\ Lett. {\bf B 686}, 44 (2010).

\bibitem{bamba} K.~Bamba, S.~Nojiri and S.~D.~Odintsov,
JCAP  {\bf 0810}, 045 (2008).

\bibitem{Bouhmadi} M.~Bouhmadi-Lopez and J.~A.~Jimenez-Madrid,
JCAP {\bf 0505}, 005 (2005).

\bibitem{Khalatnikov} I.~M.~Khalatnikov,
Phys.\ Lett.\ {\bf B 563}, 123 (2003).

\bibitem{Kamenshchik} A.~Yu.~Kamenshchik, U.~Moschella and V.~Pasquier,
Phys.\ Lett. {\bf B 511}, 265 (2001).

\bibitem{Bilic} N.~Bilic, G.~B.~Tupper and R.~D.~Viollier,
Phys.\ Lett.\ {\bf B 535}, 17 (2002).

\bibitem{Bento} M.~C.~Bento, O.~Bertolami and A.~A.~Sen,
Phys.\ Rev. {\bf D 66}, 043507 (2002).

\bibitem{Bouhmadi-2} M.~Bouhmadi-Lopez, P.~F.~Gonz$\acute{a}$lez-Di$\acute{a}$z
and P.~Martin-Moruno, gr-qc/0612135.

\bibitem{Yurov} A.~V.~Yurov, A.~V.~Astashenok, and
P.~F.~Gonz$\acute{a}$lez-Di$\acute{a}$z,
Grav.\ Cosmol. {\bf 14}, 205 (2008).

\bibitem{Yurov-4} A.~Yurov, Eur.\ Phys.\ J.\ Plus, {\bf 126} (2011) 132.

\bibitem{Nojiri-3} S.~Nojiri, S.~D.~Odintsov, and S.~Tsujikawa,
Phys. Rev. {\bf D 71}, 063004 (2005); \\
S.~Nojiri and S.~D.~Odintsov, Phys.\ Rev. {\bf D 72}, 023003 (2005).

\bibitem{barrow} J.~Barrow, Class.\ Quant.\ Grav. {\bf 21}, L79 (2004).

\bibitem{Stefanic} H.~Stefancic, Phys.\ Rev. {\bf D 71}, 084024 (2005).

%\cite{Sahni:2002dx}
\bibitem{Sahni:2002dx}
V.~Sahni and Y.~Shtanov,
%``Brane world models of dark energy,''
JCAP {\bf 0311}, 014 (2003)
[astro-ph/0202346].
%%CITATION = ASTRO-PH/0202346;%%


\bibitem{Frampton-2} P.~H.~Frampton, K.~J.~Ludwick and R.~J.~Scherrer,
arXiv:1106.4996.

\bibitem{Frampton-3} P.~H.~Frampton, K.~J.~Ludwick and R.~J.~Scherrer,
arXiv:1112.2964.

\bibitem{LR}
I.~Brevik, E.~Elizalde, S.~Nojiri and S.~D.~Odintsov, arXiv:1107.4642; \\
P.~Frampton and K.~J.~Ludwick, arXiv:1103.2480; \\
S.~Nojiri, S.~D.~Odintsov and D.~Saez-Gomez, arXiv:1108.0767; \\
L.~N.~Granda and E.~Loaiza, arXiv:1111.2454; \\
P.~Xi, X.~Zhai and X.~Li, Phys.\ Lett. {\bf B 706}, 482 (2012); \\
M.~Ivanov and A.~Toporensky, arXiv: 1112.4194; \\
%\cite{Makarenko:2012gm}
%\bibitem{Makarenko:2012gm}
%\cite{Belkacemi:2011zk}
%\bibitem{Belkacemi:2011zk}
M.~-H.~Belkacemi, M.~Bouhmadi-Lopez, A.~Errahmani and T.~Ouali,
%``The holographic induced gravity model with a Ricci dark energy: smoothing
%the little rip and big rip through Gauss-Bonnet effects?,''
arXiv:1112.5836 [gr-qc]; \\
%%CITATION = ARXIV:1112.5836;%%
A.~N.~Makarenko, V.~V.~Obukhov and I.~V.~Kirnos,
%``From Big to Little Rip in modified F(R,G) gravity,''
arXiv:1201.4742 [gr-qc].
%%CITATION = ARXIV:1201.4742;%%

\bibitem{Wei} H.~Wei and R.~G.~Cai, Phys.\ Rev. {\bf D 72}, 123507 (2005); \\
H.~Wei, Phys.\ Lett. {\bf B 695}, 307 (2011).

\bibitem{Chimento} L.~P.~Chimento, Phys.\ Rev. {\bf D 81}, 043525 (2010).

\bibitem{Cai} R.~G.~Cai and Q.~P.~Su, Phys.\ Rev. {\bf D 81}, 103514 (2010).

\bibitem{ito} Y.~Ito, S.~Nojiri and S.~D.~Odintsov, arXiv:1111.5389.

\bibitem{Kiril-Starobinsky} K.~A.~Bronnikov, M.~V.~Skvortsov and
A.~A.~Starobinsky, Grav.\ Cosmol. {\bf 16}, 216 (2010).

\bibitem{BigTrip-04} P.~F.~Gonz$\acute{a}$lez-Di$\acute{a}$z,
Phys.\ Rev.\ Lett. {\bf 93}, 071301 (2004); \\
P.~F.~Gonz$\acute{a}$lez-Di$\acute{a}$z and J.~A.~Jimenez-Madrid,
Phys.\ Lett. {\bf B596}, 16 (2004).

\bibitem{BigTrip-06} A.~V.~Yurov, P.~M.~Moruno and
P.~F.~Gonz$\acute{a}$lez-Di$\acute{a}$z, Nucl.\ Phys. {\bf B 759}, 320 (2006).

\bibitem{BigTrip-Faraoni} V.~Faraoni and W.~Israel,
Phys.\ Rev.\ {\bf D 71}, 064017 (2005).

\bibitem{BigTrip-rejoinder-1} P.~F.~Gonz$\acute{a}$lez-Di$\acute{a}$z,
Phys.\ Lett. {\bf B 632}, 159 (2006); Phys.\ Lett. {\bf B 635}, 1 (2006).

\bibitem{Bab-Doc-Ero} E.~Babichev, V.~Dokuchaev and Yu.~Eroshenko,
Phys.\ Rev.\ Lett. {\bf 93}, 021102 (2004).

\bibitem{Frampton-4} P.~H.~Frampton, K.~J.~Ludwick, S.~Nojiri, S.~D.~Odintsov
and R.~J.~Scherrer, arXiv:1108.0067.

\bibitem{w-singularity} M.~P.~Dabrowski and T.~Denkiewicz,
Phys.\ Rev.\ {\bf D79}, 063521 (2009).

\bibitem{Brane-like-Yu} A.~V.~Yurov, Phys.\ Lett. {\bf B 689}, 1 (2010).

\end{thebibliography}
\end{document}